\documentclass[12pt,aasms5]{aastex}

\def\ltsima{$\;\buildrel < \over \sim \;$}
\def\simlt{\lower.5ex \hbox{\ltsima}}
\def\gtsima{$\;\buildrel > \over \sim \;$}
\def\simgt{\lower.5ex \hbox{\gtsima}}
\begin{document}
\title{557 GHz Observations of Water Vapor Outflow from VY CMa and W Hydrae}

\author{Martin Harwit} 
\affil{511 H St., SDW, Washington, DC 20024; also Cornell University}
\and 
\author{Edwin A. Bergin}
\affil{Harvard-Smithsonian Center for Astrophysics, 60 Garden St. Cambridge, MA, 02138}

\begin{abstract}

We report the first detection of thermal water vapor emission in the 557 GHz,
$1_{10} - 1_{01}$ ground state transition of ortho-H$_2$O toward VY Canis Majoris.   In
observations obtained with the Submillimeter Wave Astronomy Satellite (SWAS), we measured
a flux of $\sim 450\,$Jy, in a spectrally resolved line centered on a velocity $v_{LSR} =
25\,$km\,s$^{-1}$ with a
full width half maximum of $\sim 35$ km\,s$^{-1}$, somewhat dependent on the assumed line
shape.  We
analyze the line shape in the context of three different radial outflow models for which we
provide analytical expressions. We also detected a weaker 557 GHz  emission line from W
Hydrae. We find that these and other H$_2$O emission line strengths scale as suggested by
Zubko and Elitzur (2000).
 
\end{abstract}

\keywords{circumstellar matter -- infrared: stars -- stars: mass loss -- stars: winds, outflows-- 
stars: individual (VY Canis Majoris, W Hydrae)}

\section{Introduction}

VY Canis Majoris is a highly luminous, variable supergiant. It is strongly obscured; only
1\,\% of the total luminosity is detected at optical wavelengths.  The  line of sight
optical depths to the star at wavelengths of 1.65, 2.26 and 3.08\,$\mu$m are 1.86 $\pm$ 0.42,
0.85 $\pm$ 0.20 and 0.44 $\pm$ 0.11 (Monnier et al. 1999). Correspondingly, the star exhibits a
huge infrared
excess. The star's distance is believed to be $D =1500$\,pc,  implying a luminosity $L =
5\times 10^5 L_{\odot}$. (cf. Lada \& Reid 1978, Richards et al. 1998). A high signal-to-noise
ratio infrared spectrum shows the star enveloped in an optically thick  dust cloud whose inner
edge is at a temperature of several hundred degrees (Harwit et al. 2001, hereafter Paper I). At
greater distances from the star the spectrum is dominated by amorphous silicate features.

Strong outflows have been reported by Buhl et al. (1975) who detected thermal emission in
the vibrational ground state (v = 0, J = 2 - 1) 86.84686 GHz SiO line.  Reid and Dickinson
(1976) interpreted these in terms of a stellar radial velocity of $17.6 \pm 1.5$\,km\,s$^{-1}$
with respect to the local standard of rest, and an expansion velocity of 36.7$\pm
2.0$\,km\,s$^{-1}$.  Three thermally emitted mid-infrared water vapor emission lines seen in
high-resolution spectra obtained with the Short Wavelength Spectrometer (SWS) on the Infrared
Space Observatory (ISO) indicate a mean radial velocity of order $20 \pm 2$\,km\,s$^{-1}$ and
a $25$\,km\,s$^{-1}$ outflow velocity (Neufeld et al., 1999).  These water vapor outflow
velocities are significantly lower than the velocities for SiO cited by Reid \& Dickinson (1976) or
the 32\,km\,s$^{-1}$ velocities that Reid \& Muhleman (1978) reported for 1612 MHz OH
maser outflow.  

The effective temperature of the central star appears to be $T_*= 2,800$\,K (Monnier
et al. 1999).  The combination of effective temperature,
distance, and inferred luminosity fixes the star's radius at $R_*\sim 15$\,AU.  In the near-infrared
at 0.8, 1.28 and 2.17\,$\mu$m, VY CMa appears embedded in an elongated circumstellar
envelope with respective dimensions of $60\times 83$, $80\times 166$ and $138\times
205$\,mas (Wittkowski et al. 1998).  Working at 11\,$\mu$m, Danchi et al. (1994) reported a
variable photospheric radius ranging from 9.5 to 11 mas, where 10 mas corresponds to a stellar
radius of 15 AU;  the surrounding dust shell had a corresponding inner radius of 40 to 50 mas
and an outer radius of 2.5 arcsec. 

To model the dust disk surrounding VY CMa, Efstathiou \& Rowan-Robinson (1990, hereafter
 ERR) assumed that the central star is orbited by a massive dusty disk, which flares out from the
star at angles $\pm 45^{\circ}$ above and below the disk's central plane.  At 
angles $> 45^{\circ}$ off the plane, radiation escapes into space. ERR believe that the line of
sight from Earth is inclined at about 43$^{\circ}$ to the disk's orbital plane, so that we are
looking almost tangentially through the outer layers of the disk.  This radiative transfer model has
been widely accepted because it accounts for the dust spectrum, the high reddening, and the lack
of visible radiation from the star.  However, as  Ivezi\'{c} \& Elitzur (1995) have
emphasized, a correct model must self-consistently couple the equations of radiative transfer to
the outflow dynamics. Zubko \& Elitzur (2000, hereafter ZE) have recently carried out such a
calculation, and used it to extensively model the semi-regular M7.5 red giant W Hydrae.

In this letter we re-examine the outflow problem.  In \S 2, we present new data on the
strength and shape of the 557 GHz water vapor line observed with the
Submillimeter Wave Astronomy Satellite (SWAS) in VY CMa and W Hydrae.   In
\S 3 we introduce complementing  ISO data.  \S 
4 attempts to model the geometry of the VY CMa outflow, while \S 5 portrays its radial
dependence. A concluding section summarizes our findings.

\section{Observations}

In the course of many  hundreds of SWAS satellite orbits, between November 1999 and May
2000, we carried out observations on VY CMa in  the 556.936 GHz $1_{10} - 1_{01}$ ground
state transition of ortho water vapor.   All observations reported here were made with the
spacecraft in the nodding mode (Melnick et al. 2000), with the telescope's primary field of view
centered on $\alpha(2000) = 07^h\ 22^m\ 58.3^s$, $\delta(2000) =$ $-25^{\circ}\  46'\ 03\farcs
0$. The observing time accumulated on source was 7902 minutes (131.7 hours) with an equal
time devoted to a blank field displaced south in declination by 0.24$^{\circ}$. 

The observed water vapor line, shown as a histogram in Figure 1, is centered at 556.89 GHz,
corresponding to a velocity with respect to the local standard of rest $v_{LSR} =
25$\,km\,s$^{-1}$.  A fit to the line shape discussed in \S 4 yields a full width at half
maximum (FWHM) of $\sim 35$\,km\,s$^{-1}$; the peak signal to noise ratio is $S/N =7$. 
The area
under the fit divided by its FWHM yields an antenna temperature $T_A^* = 0.033\pm
0.005\,{\rm K}\,\sim
450$\,Jy.  

Using a beam centered on $\alpha(2000) = 13^h\ 49^m\ 02.1^s$, $\delta(2000) = -28^{\circ}\ 22'\
03\farcs 0$, we carried out
observations on W Hydrae in January and February 1999, July 1999, July 2000,
and February 2001 for a total of 126.7 hours, and also detected the 557 GHz line.  The antenna
temperature was  $T_A^* = 0.016\pm0.004\,$\, K with a line width $\sim 10.4\pm 2.1$
km\,s$^{-1}$.

\section{Complementary Water Vapor Data}

Neufeld et al. (1999) reported line strengths for more than forty thermal water
vapor emission lines from VY CMa observed with ISO, in the
wavelength range of $\sim30$ to 45\,$\mu$m. Three transitions, respectively at 29.84, 31.77 and
40.69\,$\mu$m, were also observed with SWS in its Fabry-Perot mode and exhibited P Cygni
profiles and emission lines peaked at an
average velocity of $v_{LSR}\sim 20$\,km\,s$^{-1}$, consistent with the SWAS
observations obtained at a higher velocity resolution of $\sim 1$\,km s$^{-1}$.  The outflow
velocity of
$\sim25$\,km\,s$^{-1}$ inferred by Neufeld et al. (1999), is also in reasonable agreement with
our 557\,GHz results.  The line profile obtained with SWAS does not exhibit P Cygni
profiles, nor would we expect it to because the continuum is too weak to be detected.
Archival ISO data obtained with the Long Wavelength Spectrometer, LWS, provide additional
flux levels for transitions at 66.44, 108.078, 174.62, and 179.53$\,\mu$m.  Table 1 compares
these data on VY CMa to measurements on W Hydrae compiled by ZE and found to be in  
reasonably good agreement with their theoretical model for the star.  The table shows the flux
levels in VY CMa and W Hydrae to be approximately in the ratio 10:1.  The  VY CMa data
scaled down by a factor of 10 correspond well to the ZE model.  

\section{Modeling the VY CMa Outflow}

The outflow from red supergiants is generally assumed to be propelled by radiatively
accelerated dust grains that transfer their momentum to ambient gas.  The VY CMa mass loss
rates cited by different observers range from $\sim$\,1 to 3$\times 10^{-
4}M_{\odot}\,{\rm yr}^{-1}$ (cf. Paper I for a summary). The water vapor outflow
can be expected to be roughly 1\% of this by mass and, given the high Einstein coefficient for the
557 GHz transition, $A = 3.45\times 10^{-3}$\,s$^{-1}$, the observed emission has to be
optically thick. Velocity gradients, however, permit the radiation to escape; otherwise we would see only a velocity component from the approaching part of the outflow, rather than from its entirety.

Despite the disk-shaped outflow postulated by ERR, the structure of the cloud surrounding VY
CMa is quite uncertain.  Planetary nebulae often light up bipolar flows that may have been
ejected during a previous evolutionary phase.
Fig 2(a) provides coordinates for an axially symmetric outflow whose axis of
symmetry is inclined toward the observer at an angle $\theta_0$.  A spherically symmetric,
constant velocity, optically-thick outflow (model A) would yield a parabolic
line shape shown as the solid curve in Figure 2b.  

For the bi-conical geometry shown in Figure 2a, the flow directed at angle $\theta$ to the
observer's line of
sight is radially outward from the origin and describes a circular arc of radius $aa' = aa''$.  Its
length $\ell$ spans points $a'$ and $a''$ on a circle defined by the angle $\theta_0$, by the
opening angle $\alpha$ of the cone, and by the tilt angle toward the observer, $\theta$, of the
conical surface joining the origin to points on the circular arc.  The length of the arc, $\ell$ can
be expressed as
\begin{equation}
\ell = R \sin \theta \biggl [ \pi - 2\sin^{-1}\biggl ( \frac{\sin (\theta_0 - \alpha) + [\cos(\theta_0 -
\alpha) -\cos\theta]\cot\theta_0}{\sin\theta}\biggr ) \biggr ]
\end{equation}

The segment formed by joining the origin to points along this arc presents to the observer a 
pie-shaped area projected on the plane of the sky, of measure  $(R\ell\sin\theta)/2$ defined by
points $a$, $a'$, and $a''$.  The projected velocity along the observer's line of sight throughout
this surface is $v = v_0 \cos\theta$; for an optically thick line the velocity distribution yielding
the line shape is given by the relative values of $(R\ell\sin\theta)/2$ at different angles $\theta =
\cos^{-1}(v/v_0)$. 

An optically-thick, bi-conical, constant-velocity outflow, symmetric about an
axis perpendicular to the observer's line of sight (i.e. with $\theta_0 = \pi /2$) and  limited to
angles less than $\alpha =\pi /3$ from the axis of symmetry --- with a density independent of
angular distance from the axis (model B),  yields a line profile shown by the dashed curve in
Figure 2b.  A uniform bi-conical outflow within an angle of $\alpha = \pi/4$ from the axis, and
with $\theta_0 = 47^{\circ}$ (model C), would yield the bi-lobed dotted line shape shown in
Figure 2b.  Neither of these two models produces the line shape observed.

To obtain the line profile for an optically-thick disk, we subtract these respective
bi-conical outflows
from a spherical outflow. For a disk with axis perpendicular to the line of sight to Earth, and a
latitude-independent outflow into a disk with half angle $\pm 30^{\circ}$ above and below the
symmetry plane, we can subtract the dashed curve in Fig 2b from the solid curve --- i.e. models
(A - B) --- to obtain the fit shown by the dashed line in Fig 1.  The line shape expected from the
model proposed by ERR $\equiv {\rm models\ (A-C)}$, would result from a subtraction of the
dotted line in Fig. 2b from the solid curve.  Figure 1 shows that models A and (A-B), normalized
to the line strength and width, give rather better fits to the observed 557 GHz profile than does
the ERR model (A-C).

\section{The Temperature/Radius Profile}

Zubko and Elitzur (2000) have devised the most comprehensive computational model of outflow
from highly evolved stars developed to date.  It derives a profile for  temperature $T$ and radial
distance $R$ from the star, $T^2R\sim$\,constant.  The gas in the outflow is primarily heated through collisions with grains, cooled by H$_2$O emission and adiabatic expansion, and effectively acts as though it
was in thermal balance with the parent star throughout the flow. In apparent agreement with the
general scaling properties proposed by Ivezi\'{c} \& Elitzur (1997) as applied to dusty gaseous
outflows, Table 1 shows the good fit of the ZE model to the observed H$_2$O line strengths in
W Hydrae, that also scales well to VY CMa.

Given this observed agreement, we adopt the ZE profile, $T^2R =$ constant, for VY CMa.  
Paper I suggests the existence of a dust photosphere at a temperature $T_p\sim 400$\,K in
thermal equilibrium with the star. $R_p$ is a characteristic radius for this photosphere in thermal
balance with the star, $R_p^2T_p^4 = R_*^2T_*^4$, so that the ZE profile can be taken to
extend all the way to the stellar surface: $T^2R = T_*^2R_*$.  In Table 3, we list the radial
distances at which the VY CMa water vapor lines emanate, on the assumption that the emitting
region is optically thick -- an assumption justified by the deep P Cyni profiles of even the highest rotational lines observed.  The flux received from a circumstellar cloud
subtending a solid angle $\Omega$ at the observer is then
\begin{equation} 
F(\nu) =\frac{2\nu^2kT}{c^2}\Omega = \frac{2kT}{\lambda^2}\frac{R^2}{D^2}\psi
\end{equation}
where we assume that all the radiation is in the Rayleigh Jeans part of the spectrum.   Table 3
shows this assumption to be consistent;  $D$ is the distance to the star, $R$ is the radial distance
of the emitting
H$_2$O region from the star, and $\psi$ is a constant of order unity that depends on the
geometry of the outflow.  For a sphere it has  a  higher value than for a flattened disk.  Table 3
assumes $\psi \sim 1$.  Substituting for the stellar parameters $T_* = 2800$\,K, $R_* =
2.25\times 10^{14}\,$cm, $D= 4.5\times 10^{21}$\,cm and taking the emission to be blackbody
in the frequency interval observed, we can solve for the characteristic distances from the central
star and the corresponding temperatures at which the different spectral lines appear to be
emitted.
\begin{equation}
R = \biggl [\frac{1.75\times 10^{25}F(\nu)\lambda^2}{\psi}\biggr ]^{2/3}
\end{equation}
where $F(\nu)$ is given in Janskys, and where $R$ and $\lambda$ both are in units of
centimeters.  The calculated distances for the last three lines in Table 3 use the full emission line
strengths corrected for P Cygni absorption of the continuum.  For the longer wavelength lines
observed with the ISO LWS, the resolution is insufficient to determine whether self-absorption
weakened the lines.

\section{Conclusions}

We have obtained a high-resolution spectrum of  the 557 GHz water vapor emission line from
the circumstellar cloud surrounding VY CMa, and have obtained a weaker signal from W
Hydrae.  The observed flux for VY CMa  is $\sim 450\,$Jy.  The line shape
is symmetric around a recession velocity $v_{LSR} = 25\,$km\,s$^{-1}$ and has a FWHM of
order $\sim 35$\,km\,s$^{-1}$.  We have developed analytical line fits for several plausible models of
the outflow. 
Observations that will be possible with the Herschel space telescope should yield improved
line shapes that will further constrain outflow models.  Archival and published ISO data for VY
CMa and W Hydrae, suggest the adoption of a model developed by Zubko \& Elitzur (2000) to
determine the distance from the central star and the temperature regime around VY CMa from
which the H$_2$O lines emanate.

\section{Acknowledgments}

We have been supported by contract NAS-30702 from NASA for work
on the Submillimeter Wave Astronomy Satellite, SWAS, and contract NASA/JPL 1203030 for
work on the Infrared Space Observatory, ISO.  We thank Gary Melnick and David Neufeld for critically reading our manuscript, the referee Keiichi Ohnaka for thoughtful and incisive comments, and Viktor Zubko and Moshe Elitzur for their results of unpublished calculations.
\centerline{\bf References}
\vskip 0.1 true in 
{\hoffset 20pt
\parindent = -20pt

Barlow, M. J., et al. 1996, A\&A 315, L241

Buhl, D., Snyder, L. E., Lovas, F. J. \& Jonson, D. R. 1975, ApJ 201, L29

Danchi, W.C., Bester, M., Degiacomi, C.G., Greenhill, L.J. \& Townes, C.H. 1994, AJ 107, 1469

Efstathiou, A. \& Rowan-Robinson, M. 1990, MNRAS 245, 275 (ERR)

Harwit, M., Malfait, K., Decin, L., Waelkens, C., Feuchtgruber, H., \& Melnick, G., J., 2001, ApJ 557, 844

Ivezi\'c, \v{Z}. \& Elitzur, M. 1995, ApJ 445, 415

Ivezi\'c, \v{Z}. \& Elitzur, M. 1997, MNRAS 2287, 799

Lada, C.J. \& Reid, M.J. 1978, ApJ 219, 95

Melnick, G. J. et al. 2000, ApJ 539, L77

Monnier, J.D., Geballe, T.R. \& Danchi, W.C. 1999, ApJ 512, 351

Neufeld, D. A. et al. 1996, A\&A 315, L237

Neufeld, D.A., Feuchtgruber, H., Harwit, M. \& Melnick, G. 1999, ApJ 517, L147

Reid, M. J. \& Dickinson, D. F. 1976, ApJ 209, 505.

Richards, A.M.S., Yates, J.A. \& Cohen, R.J. 1998, MNRAS 299, 319

Wittkowski, M., Langer, N. \& Weigelt, G. 1998, A\&A 340, L39

Zubko, V. \& Elitzur, M. 2000, ApJ 544, L137 (ZE)

}

\vfill\eject
\centerline{\bf Figure Captions}
\vskip 0.1 true in 
{\hoffset -20pt
\parindent -20pt

Figure 1.  The 557\,GHz water vapor spectrum of VY CMa obtained with SWAS.  The histogram
gives the SWAS observations.   The solid  line is a fit for a spherically symmetric outflow (model
A). The dashed line is for a disk viewed edge-on (models A - B); the dotted curve is a fit for the
ERR outflow (models A - C).  All three curves are normalized to the observed line profile (see
text and Figure 2 for details).

Figure 2. (a) The geometry of the outflow for which the velocity profiles were derived.  $R$ is
the
radius of a sphere whose radial extent is that of the outflow.  For a bi-conical or disk-shaped
flow, the angle $\theta_0$ gives the tilt angle toward the line of sight. Throughout, we assume
for simplicity that the symmetry axis and the line of sight define a plane of mirror symmetry in 
the outflow.  The angle $\alpha$ is the opening angle of the conically shaped flow as
measured from the symmetry axis. The angle $\theta$ defines the radial direction of motion of a
volume in the flow relative to the line of sight to the observer. Points $a$, $a'$, and $a''$ lie in 
the plane of the sky.  (b) The solid curve gives the
expected line shape from a spherical outflow (model A).  The dashed curve gives the expected
line shape for a bi-conical outflow (model B) with $\theta_0 = 90^{\circ}$ and $\alpha =
60^{\circ}$, while the dotted line (model C) gives the line shape for an inclined bi-conical
outflow with $\theta_0 = 47^{\circ}$ and $\alpha = 45^{\circ}$.  Subtracting the bi-conical
outflows from that for spherical outflows, respectively, yields the dashed and dotted line profiles
for disk-shaped outflows in Figure 1. (See text for details).

}

\begin{center}
\scriptsize
\begin{tabular}{r|c|r|r|c}
\hline \multicolumn{5}{c}{Table 1. Relative o-H$_2$O Flux Levels from VY CMa and W
Hydrae}\\
\hline

Wavelength & Transition  & W Hydrae$^a$ & Model$^b$ & VY CMa$^c$ \\
$\lambda$\ ($\mu$m) & & $10^{-20}$W\,cm$^{-2}$ & $10^{-20}$W\,cm$^{-2}$ &
$10^{-19}$W\,cm$^{-2}$\\  
\hline           
538.289 & $1_{10}\rightarrow 1_{01}$ & 0.45& 0.76 & 0.34\\
179.527 & $2_{12}\rightarrow 1_{01}$ & 8.66 & 12.48 & 8.7\\
174.624 & $3_{03}\rightarrow 2_{12}$ & 9.21 & 8.31 & 8.8\\
108.073 & $2_{12}\rightarrow 1_{10}$ & 13.00 & 15.47 & 13.2\\
66.438 & $3_{30}\rightarrow 2_{21}$ & 22.90 & 18.45 & 17.2\\
40.691$^d$ & $4_{32}\rightarrow 3_{03}$ & 23.00 &36.07 & 36.9\\
31.772 $^d$& $4_{41}\rightarrow 3_{12}$ & 63.0 & 39.95 & 24.1\\
29.837 & $7_{25}\rightarrow 6_{16}$ & 32.00 & 30.72 & 23.5\\
\hline
\multicolumn{5}{l} {$^a$ ISO from Neufeld et al. (1996), Barlow et al. (1996); SWAS data,
this paper}\\ 
\multicolumn{5}{l}{$^b$ W Hydrae model from ZE, and 538\,$\mu$m flux through}\\
\multicolumn{5}{l}{private communication from Drs. Zubko \& Elitzur}\\
\multicolumn{5}{l} {$^c$ For sources of data see text}\\  
\multicolumn{5}{l}{$^d$ Blending may need to be taken into account}\\
\hline
\end{tabular}
\end{center}
\normalsize

\begin{center}
\scriptsize
\begin{tabular}{c|c|c}
\hline \multicolumn{3}{c}{Table 2. Characteristics of VY CMa and W Hydrae}\\
\hline

Star   &  VY CMa $^a$& W Hydrae$^b$ \\
Star's Distance (pc) & 1500 & 115\\
Star's Luminosity ($L_{\odot}$) & $5\times 10^5$ & 11,050\\
Star's Temperature (K) & 2800 & 2500\\
Star's Radius (cm) & $2.25\times 10^{14}$ & $3.9\times 10^{13}$\\
Final Outflow Velocity (km s$^{-1}$) & 20 & 10\\
Optical Depth & $A_J \sim 3.2$ & 0.83 at 5500\AA\\
Gas-to-dust ratio & $\sim 100$ & 850\\
Mass loss rate $M_{\odot}$ yr$^{-1}$ & $\sim 2\times 10^{-4}$ & $2.3\times 10^{-6}$\\
\hline
\multicolumn{3}{l}{$^a$ Data from Neufeld et al. (1996), Barlow et al. (1996)}\\
\multicolumn{3}{l} {$^b$ For sources of data see text}\\
\hline
\end{tabular}
\end{center}
\normalsize

\begin{center}
\scriptsize
\begin{tabular}{r|r|r|r|r|r}
\hline \multicolumn{6}{c}{Table 3. Ortho-H$_2$O Emission: Characteristic Distances and
Temperatures}\\
\hline
Wavelength & Flux & Radius$^a$ & Temperature$^a$ & $T_u\ ^b$ & $T_l\ ^b$ \\
($\mu$m) &  (Jy) & ($10^{16}$\,cm) & K & K & K\\
\hline           
538.289 & 450 & 8.1 & 148 & 60.95 & 34.23\\
179.527 & 3490 & 7.3 & 155 & 114.35 & 34.23\\
174.624 & 3440 & 7.2& 156 & 196.72 & 114.35 \\
108.073 & 3190 & 6.9 &160 & 194.05 & 60.95\\
66.438 & 2560 & 1.6 &332 & 410.55 & 194.05\\
40.691$^c$ & 3360 & 0.99 &431 & 550.30 & 196.72 \\
31.772$^c$ &1710& 0.45  & 626 & 702.21 & 249.37\\
29.837 & 1570 & 0.39  & 672 & 1125.55 & 643.34\\
\hline
\multicolumn{6}{l} {$^a$ Derived from constancy of $TR^{1/2}$, see text}\\ 
\multicolumn{6}{l}{$^b$ Level energies divided by Boltzmann constant}\\
\multicolumn{6}{l}{$^c$ Blending may need to be taken into account}\\

\hline
\end{tabular}
\end{center}
\normalsize

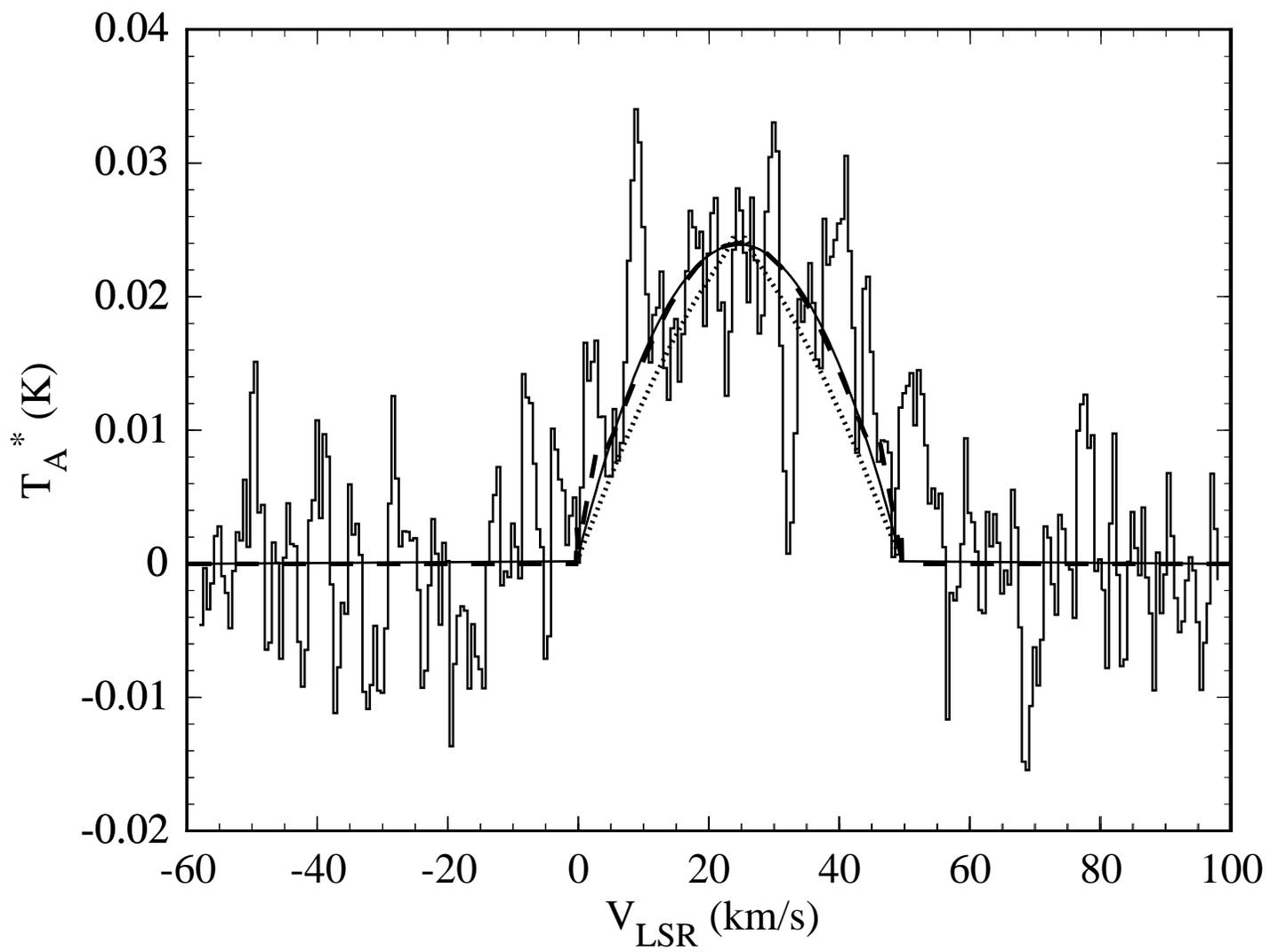
\end{document}